\documentclass[usegraphicx,usenatbib,a4paper]{mn2e}
\usepackage{aas_macros,graphicx,times,multirow}
\usepackage{amssymb}
\usepackage{float}

\title[On the nature of SWIFT\,J1907.3-2050, IGR\,J12123-5802, and IGR\,J19552+0044]
{On the nature of the hard X-ray sources SWIFT\,J1907.3-2050, 
IGR\,J12123-5802 and IGR\,J19552+0044}
\author[F. Bernardini et al.]                                                    
{F.~Bernardini,$^{1,2}$\thanks{E-mail: bernardini@wayne.edu} 
D.~de Martino$^{2}$, 
K.~Mukai$^{3,4}$, 
M.~Falanga$^{5}$, 
I.~Andruchow$^{6}$,
\newauthor
J.-M.~Bonnet-Bidaud$^7$,
N.~Masetti$^{8}$,
D.H.~Gonzalez Buitrago$^9$, 
M.~Mouchet$^{10,11}$,
G.~Tovmassian$^{9}$\\
$^1$ Wayne State University, 666 W. Hancock Street, Detroit, MI, USA\\
$^2$ INAF $-$ Osservatorio Astronomico di Capodimonte, Salita Moiariello 16, I-80131 Napoli, Italy\\
$^3$  CRESST and X-Ray Astrophysics Laboratory, NASA Goddard Space Flight  Center, Greenbelt, MD 20771, USA\\
$^4$ Department of Physics, University of Maryland, Baltimore County, 1000 Hilltop Circle,Baltimore, MD 21250, USA\\
$^5$ International Space Science Institute (ISSI), Hallerstrasse 6,CH-3012 Bern, Switzerland \\
$^6$  Facultad de Ciencias Astronomicas y Geofisicas, UNLP, and Instituto de
Astrofisica La Plata, CONICET/UNLP, Argentina \\
$^7$ CEA Saclay,  DSM/Irfu/Service d'Astrophysique, F-91191, Gif-sur-Yvette, France \\
$^8$ INAF $-$ Istituto Astrofisica Spaziale, Via Gobetti 101, I-40129, Bologna, Italy \\
$^{9}$ Instituto de Astronomia, Universidad Nacional Autonoma de Mexico, Apdo. Postal 877, Ensenada, 
Baja California 22800, Mexico\\
$^{10}$ Laboratoire APC, Universit\'{e} Denis Diderot, 10 rue Alice Domon et L\'{e}onie Duquet, F-75005
Paris, France \\ 
$^{11}$ LUTH, Observatoire de Paris, Section de Meudon, 5 place Jules Janssen, F-92195
Meudon, France}

\date{Accepted 2013 July 30.  Received 2013 July 25; in original form 2013 June 19}
\def\Swift{{\em Swift}}

\def\XMM{{\em XMM-Newton}}
\def\Suzaku{{\em Suzaku}}

\def\ergscm{$\rm erg\,cm^{-2}\,s^{-1}$}
\def\INT{{\em INTEGRAL}\,}

\def\JA{\hbox{Swift\,J1907}}
\def\JB{\hbox{IGR\,J1212}}
\def\JC{\hbox{IGR\,J1955}}
\def\JAA{\hbox{Swift\,J1907.3-2050}}
\def\JBB{\hbox{IGR\,J12123-5802}}
\def\JCC{\hbox{IGR\,J19552+0044}}

\begin{document}

\label{firstpage}
\maketitle
\begin{abstract}

 The \INT\ and \Swift\ hard X-ray surveys have identified a large number of new sources, among which many are 
proposed as Cataclysmic Variables (CVs). Here we present the first detailed study of three X-ray selected CVs, 
\JAA, \JBB, and \JCC\  based on \XMM, \Suzaku, \Swift\ observations and ground based optical and archival 
nIR/IR data.  \JAA\ is highly variable  from hours to months-years at all  wavelengths. No coherent 
X-ray pulses are detected but rather transient features.  The X-ray spectrum reveals a multi-temperature 
optically thin plasma absorbed by complex neutral material and a soft black body component arising from a small 
area. These characteristics are remarkably similar to those observed in magnetic CVs. A supra-solar abundance of 
nitrogen could arise  from nuclear processed material from the donor star.  \JAA\ could be a peculiar 
magnetic CV with the second longest (20.82\,h) binary period. \JBB\ is  variable in the X-rays on a 
timescale of $\gtrsim 7.6$ h. No coherent pulsations are detected, but its spectral characteristics 
suggest that 
it could be a magnetic CV of the Intermediate Polar (IP) type.  \JCC\ shows two X-ray periods, $\sim1.38$ 
h and $\sim1.69$ h and a X-ray spectrum  characterized by a multi-temperature plasma with little 
absorption.   We derive a low accretion rate, consistent with  a CV below the orbital period gap. Its  
peculiar nIR/IR spectrum suggests a contribution from cyclotron  emission. It could either be a 
pre-polar or an IP with the lowest degree of asynchronism.

\end{abstract}
\begin{keywords}
Cataclysmic Variables: general -- stars: withe dwarf -- X-rays: individual: Swift J1907.3-2050 (also 
known as: V1082 Sgr), Swift J1212.3-5806  (also known as: IGR J12123-5802 and 1RXS 
J121222.7-580118), IGR J19552+0044, WISE\,J195512.47+004536.6
 \end{keywords}

\section{Introduction}
Cataclysmic Variable stars (CVs) are binary system in which a   white dwarf (WD) accretes matter from a 
late-type main 
sequence or subgiant star. 
 A large fraction,  $\sim20-25\%$ \citep[see e.g.][and reference therein]{wickramasinghe00,pretorius13} of whole CV class, 
harbours magnetized WDs with B$\gtrsim10^{5}$ G up to $2.3\times10^{8}$ G. 
The high magnetic systems, the Polars, show intense and variable polarized emission in the optical and near-IR ranges,
while those showing weakly or unpolarized optical/nIR emission, the Intermediate Polars (IPs), are believed to possess weaker B$\leq10^{6}$ G. They 
are both characterized by variable X-ray emission at the rotational period of the accreting WD  
that is $\rm P_{rot}/P_{orb}\simeq1$ in the Polars (synchronous  systems) and $\rm P_{rot}/P_{orb}<1$ in the IPs 
(asynchronous  systems). 
The latter group is generally found above the 2--3 h orbital period gap, whilst the Polars populate the orbital period 
distribution at short values, mainly below the gap \citep[see][]{Warner95}. 

\noindent The hard X-ray surveys carried out above 20 keV by the \INT\ and \Swift\ satellites have detected more than a 
thousand 
 of sources, a non-negligible fraction ($\sim6\%$) of them are galactic X-ray binaries containing WDs 
 \citep{barlow06,bird10,Cusumano10,Baumgartner13}. 
The confirmed CVs detected in these surveys are mainly magnetic of the IP type ($\sim 80\%$ of the whole sample). 
A few  Polars and a handful of non-magnetic systems that include Dwarf Novae, old Novae and nova-like CVs,  the 
latter still disputed to be magnetic, are also present. Finally, a few symbiotic stars have been also detected \citep[see e.g.][]{luna12}. 
The current  roster of these hard X-ray accreting WDs totals to 67, about half of them have been first proposed as CV 
candidates  through  optical spectroscopic follow-ups \citep[e.g.][]{masetti06,masetti08,Masetti10,Masetti12a} and many are 
suspected to harbour a magnetic WD. 
However, only X-ray follow-ups, mainly conducted with \XMM, allowed to unambiguously confirm 
\citep{demartino08,Anzolin09,scaringi11,Bernardini12,Masetti12b} or reject 
\citep{deMartino10,deMartino13} a CV identification, thanks to the detection (or non detection) of an X-ray signal at the WD spin 
period  and to the characterization of the X-ray spectral properties. However, so far, there are 
several tens of sources to be still identified.

 \noindent The role of CVs in the galactic population of X-ray sources has been quite discussed recently. In particular, 
 magnetic CVs, are claimed to represent a large population (thousands of sources), and to be dominant 
contributors to the 
 Galactic Ridge X-ray Emission (GRXE) and bulge emission \citep{revnivtsev11,hong2012a,hong2012b} at low luminosities 
 (i.e. below $\rm 10^{33}\,erg\,s^{-1}$). However, the bulk of the CV population that is dominated by 
non-magnetic short orbital period systems, can account  for the very low end ($\rm < 10^{30}\,erg\,s^{-1}$) of 
the GRXE luminosity function \citep{Reis13}, while magnetic CVs are believed to contribute most at $\rm \gtrsim 
10^{30}\,erg\,s^{-1}$ \citep{hong2012b}.  The contribution from X-ray active stars at very low luminosities is still debated. 
The ratio of WD binaries and of coronally active stars is believed to be $\sim$2:1 
 \citep{Morihana13,nebot13}.

 \noindent In magnetic CVs, the accretion flow is magnetically channeled  
 onto the WD polar regions. A strong shock is formed, below which matter cools via thermal 
 bremsstrahlung and cyclotron radiation, the relative proportion primarily depending on the WD magnetic field strength 
 \citep{woelk_beuermann96,fischer_beuermann01}. Hence, low-field magnetic CVs, like the IPs, have bremsstrahlung 
dominated shocks, while the Polars mainly cool via cyclotron radiation. Thus, IPs were initially known to be bright harder 
X-ray sources than the Polars. However, before the deep surveys by \INT\ and \Swift\ they only constituted a minor fraction of the 
magnetic CV subclass. This view has now changed with the identification of new magnetic systems of this type, 
suggesting that our knowledge of the magnetic CV population, and hence evolution, is still poor.

 \noindent The identification of new hard X-ray CVs, can allow to
 put constraints on the true contribution of these binaries to the whole CV population as well as to 
 identify new unexplored properties.\\ 
 Most of the \INT\ and \Swift\ hard X-ray CVs are found at long orbital periods 
($\gtrsim5$ h), with a few exceptions, and possess rapidly rotating WDs with spin-to-orbit period ratios $\leq0.12$ 
\citep{Bernardini12}. This view could be confirmed or changed by the characterization of a larger sample, so 
that the role of system parameters, such as the WD magnetic field, the WD mass and the mass accretion rate, can be 
finally unveiled. In this work, we present the first X-ray observations of a sample of three 
magnetic CV candidates, all hard X-ray sources: \JAA\ , \JBB\ and \JCC\ (henceforth \JA, \JB, and \JC). We complement 
the X-ray study with simultaneous UV and optical photometry acquired with the \XMM\ satellite and with ground based optical and archival nIR/IR data.
 \JA\ is also known as V1082\,Sgr, a long (20.82 h) orbital period CV suggested as a possible magnetic system by 
\cite{Thorstensen2010} because of the hard X-ray detection.  \JB\ and \JC\ were identified by \cite{Masetti10} from optical spectroscopic 
follow-ups as CVs and also
suspected to be magnetic because of characteristics of their emission lines. No further information on these sources are available 
to date. Our study reveals unusual temporal and spectral properties that make these sources interesting cases for further 
follow-ups at other wavelengths.

\section{Observations and data analysis}
\label{sec:obs}

\subsection{\textit{XMM-Newton} observations}
 
The three sources were all observed in 2012 by the  \XMM\ observatory, \citep{turner01,mason01,denherder01}, 
using the EPIC Cameras as main instruments. The details of the \XMM\ observations, 
together with that of \Swift\ and \Suzaku\ are reported in Tab. \ref{tab:observ}.
  All data were processed and scientific products were extracted with SAS version 12.0.1 using the latest calibration 
files (CCF) available in September 2012 (EPIC and OM) and May 2013 (RGS).

\subsubsection{The EPIC and RGS data}
 All observations were performed with the EPIC PN and MOS cameras set in 
{\tt prime full window} imaging mode with the thin filter.  Standard data screening criteria were 
  applied for all instruments.   

\noindent For EPIC data, we extracted source photons from circular regions of radius 35\arcsec\ centred on source position 
determined by Gaussian fitting on one$-$dimension photon distribution. The background photons were extracted 
from a circular regions of same size within the same CCD. 

  To avoid  contamination from solar flares in the spectral analysis we conservatively produced spectra using only those 
parts of the observations free from high background epochs.
For the timing analysis  we instead used the whole data sets except \JA\ for which contamination from solar
flares is particularly strong at the end of the EPIC exposure. Background subtracted light curves were produced in the
ranges: 0.3--15 keV  (with a bin time of 15 s),  0.3--1 keV, 1--3 keV, 3--5 keV and 5--15 keV (with a bin time of 75 s). 
For each source, event arrival times were corrected to solar system barycenter. 
The EPIC spectra were rebinned before fitting, to have at least 30 counts per bin. 
 We  report the spectral analysis results obtained with the PN data only (consistency with the results of 
 MOS cameras was always verified). Phase--resolved spectra were also extracted at the pulse maximum and 
 minimum.

\noindent  The RGS spectra were extracted using the whole exposures except \JA\ for which only the 
first 
 21\,ks of data was retained.  Due to low count rates, we only inspected the 1st order spectra for all 3 sources, however, 
 for \JB, even the 1st order spectra were too noisy to allow a spectral study.
 For \JA\ and \JC, we grouped channels so each had at least 16 source  COUNTS, after ignoring extreme ends of the RGS 
 energy ranges and
the ranges corresponding to the inoperative CCD chips.  Given the low
statistical quality and the limited energy range of the RGS data, we used
them mainly to confirm and refine the best-fit model derived using the EPIC data.

\noindent All spectra were analysed using the version of $XSPEC$ (12.7.1n).

\begin{table*}
\caption{Summary of main observations parameters for all instruments.}
\begin{center}
\begin{tabular}{cccccccc}
\hline
 & & & & & & & \\
 
Source & Telescope      & OBSID & Instrument & Date                    & UT$_{\rm start}$ & T$_{expo}$  &Net Count Rate\\
              &                          &              &        & yyyy-mm-dd      & hh:mm & (ks)$^{*}$ &    c/s                  \\
\hline
\JAA\ & \emph{XMM-Newton}&  0671850301 & EPIC-pn   & 2012-03-19  & 11:32 & 38.0/13.6$^{**}$ & $4.54\pm0.02$ \\
     &                  &              & EPIC-MOS1 & 2012-03-19  & 11:10 & 27.2 & $1.10\pm0.07$     \\ 
     &                  &              & EPIC-MOS2 & 2012-03-19  & 11:10 & 27.2 & $1.16\pm0.07$     \\  
     &                  &              & RGS1 	  & 2012-03-19  & 11:10 & 36.5/20.9$^{***}$ & $0.082\pm0.003$ \\
     &                  &              & RGS2 	  & 2012-03-19  & 11:10 & 36.5/20.9$^{***}$ & $0.093\pm0.003$  \\
     &                  &              & OM-B 	  & 2012-03-19  & 11:20 & 18.8 & $79.82\pm0.02$  \\
     &                  &              & OM-UVM2   & 2012-03-19  & 17:00 & 18.8 & $6.52\pm0.04$\\
     & \emph{Suzaku}   &  406042010        & XIS0   & 2012-03-23 & 05:32 & 39.5     & $0.692\pm0.004$    \\
     & \emph{Suzaku}   &                   & XIS1   & 2012-03-23 & 05:32 & 39.5     & $0.796\pm0.005$   \\
     & \emph{Suzaku}   &                   & XIS3   & 2012-03-23 & 05:32 & 39.5     & $0.750\pm0.005$    \\

 &  \emph{Swift} & $^{****}$ & BAT &  &	& 5800 & $3.5\pm0.4\times 10^{-4}$ \\ 
 &               & 00037329001 & XRT & 2008-04-10 & 13:38 & 3.5 & $0.25\pm0.01$  \\
 &               & 00037329002 & XRT & 2008-06-04 & 15:57 & 0.4 & $0.13\pm0.02$  \\
 &               & 00037329003 & XRT & 2008-06-12 & 13:23 & 1.0 & $0.23\pm0.02$  \\
 &               & 00037329004 & XRT & 2008-06-12 & 15:01 & 0.3 & $0.23\pm0.02$  \\
 &               & 00037329005 & XRT & 2008-06-13 & 08:23 & 0.4 & $0.36\pm0.03$  \\
 &               & 00037329006 & XRT & 2008-06-13 & 10:20 & 0.1 & $0.36\pm0.01$  \\
 &               & 00037329008 & XRT & 2008-06-26 & 00:31 & 1.4 & $0.10\pm0.01$  \\
 &               & 00031252001 & XRT & 2008-08-19 & 06:54 & 1.3 & $0.05\pm0.01$  \\
 &               & 00031252002 & XRT & 2008-08-20 & 11:41 &10.1 & $0.030\pm0.003$  \\
 &               & 00031252003 & XRT & 2008-08-28 & 07:36 & 1.9 & $0.05\pm0.01$  \\
 &               & 00037329009 & XRT & 2008-11-14 & 12:01 & 5.2 & $0.017\pm0.002$  \\
 &               & 00031252004 & XRT & 2012-06-19 & 12:32 & 3.4 & $0.002\pm0.001$ \\
 &               & 00031252005 & XRT & 2012-06-20 & 01:33 & 6.4 & $0.002\pm0.001$ \\
     &                  &             &          &             &       &       &     \\

\JBB\ & \emph{XMM-Newton}& 0671850601 &  EPIC-pn & 2012-01-07 & 15:48 & 37.4/12$^{**}$ & $0.754\pm0.008$ \\
     &                  &            &  EPIC-MOS1 & 2012-01-07 & 15:26  & 39.6 & $0.255\pm0.003$ \\
     &                  &            &  EPIC-MOS2 & 2012-01-07 & 15:26  & 39.6 & $0.259\pm0.003$ \\     
     &                  &            &    RGS1  & 2012-01-07 &  15:25 & 40.1  & $0.010\pm0.002$ \\
     &                  &            &    RGS2  & 2012-01-07 &  15:25 & 40.1  & $0.012\pm0.002$ \\
     &                  &           & OM - V   & 2012-01-07 & 15:35 &   15.8 & 2.33$\pm$0.03 \\
     &                  &           & OM - UVM2  & 2012-01-07 & 21:38 &  15.8 & 0.27$\pm$0.2 \\ 
           &  \emph{Swift} & $^{****}$ & BAT &  & & 7500 &  $1.8\pm0.3\times10^{-4}$ \\
  & & & & & & & \\
\JCC\ & \emph{XMM-Newton}& 0671850201  &EPIC-pn & 2012-04-29 & 06:04 & 38/19$^{**}$ & $2.88\pm0.02$\\
     &                  &             & EPIC-MOS1 & 2012-04-29 & 05:41 & 39 & $0.647\pm0.005$  \\
     &                  &             & EPIC-MOS2 & 2012-04-29 & 05:41 & 39 & $0.630\pm0.005 $ \\     
     &                  &             &  RGS1     & 2012-04-29 & 05:41 &  36.8  &  $0.054\pm0.002$ \\
     &                  &             &  RGS2     & 2012-04-29 & 05:41 &  36.8  &  $0.071\pm0.002$ \\
     &                  &             & OM-B     & 2012-04-29 & 05:50 &  15.8 &  4.87$\pm$0.02 \\
     &                  &             &OM-UVM2  & 2012-04-29 & 10:40 &  15.8 &  0.65$\pm$0.02 \\
\hline
\end{tabular}
\label{tab:observ}
\end{center}
\begin{flushleft}
$^{*}$ Net exposure times.\\
$^{**}$ We report both the total exposure time and that after removing solar flares.\\
$^{***}$ We report both the total exposure and that during the low background interval.\\
$^{****}$ All available pointings are summed together.\\
\end{flushleft}
\end{table*}

\subsubsection{The Optical Monitor photometry}

The Optical Monitor (OM) instrument was operated in fast window mode during all observations of the three sources.
The OM data of \JA\ and of \JC\ 
were collected in two photometric bands using first the B filter, centered at 4500\,\AA\ and then
the UVM2 filter, centered at 2310\,\AA. For \JB\ 
we instead used the V filter, centered at 5430 \AA, due to brightness limits of a close-by 
star in the OM field of view.  In this way we
obtained two time series sets for each source 18.8 ks (\JA\ ), 
 16.6\,ks (\JB\ ) and 15.8\,ks  (\JC\ ) long. 
 The OM light curves were obtained from the SAS processing pipeline with a binning time of 10\,s or 20\,s. 
 Corrections to the solar barycenter were also applied to the OM light curves.

\subsection{The \Suzaku\ observation}

Swift\,J1907 was also observed with \Suzaku\ \citep{Suzaku} in 2012 March, a few days
after the \XMM\ observation. Here we include the timing analysis of the data
taken with the X-ray imaging spectrometer (XIS; \citealt{XIS}).  We have taken
the data as processed using pipeline version 2.7.16.32, and extracted the source
light curves from a circular region of 3.50$'$ radius centered on the source
image.  Background light curves were extracted from an annular region with
outer and inner radii of 6.25$'$ and 4.50$'$, respectively, and subtracted from
the source region curves after scaling by the ratio of the extraction regions
(0.65).  Light curves from all 3 active XIS units have been combined in our
analysis.

\subsection{The \Swift\ observations}

The Swift Burst Alert Telescope, BAT \citep{Barthelmy}, is a wide-field ($\sim$1 steradian) coded aperture mask 
instrument sensitive in the 14--195 keV range.  Thanks to the large field of view, 
BAT has built up a sensitive all-sky map of the hard X-ray sky.  
We have taken the 8-channel spectra from the first 58-month of the
mission\footnote{http://swift.gsfc.nasa.gov/docs/swift/results/bs58mon/}.
We collected BAT data of \JA\ and \JB, while no data of \JC\ are available. 
\JC\ is also too faint to use the \INT\ IBIS/ISGRI spectral data,  
the source being at an average rate of 0.06$\pm$0.02 c/s in the 20-40 keV\footnote{http://www.isdc.unige.ch/heavens/}. 
Therefore, we have extended the spectral analysis above 10\,keV for \JA\ and \JB\ only.

\noindent The \emph{Swift} X-ray Telescope \citep[XRT,][]{Gehrels_2004} 
is an imaging CCD spectrometer sensitive in the 0.3--10 keV range. 
\emph{Swift} performed 13 short (few ks) pointings of \JA, 11 in 2008 and 2 in 2012.  
A total of about 35 ks of exposure time was collected on this source.
We used the \emph{Swift}/XRT data products generator at the University of
Leicester \citep{Evans09} to build the background subtracted light curve and the average spectrum in photon 
counting (PC) mode of \JA.

\begin{figure}
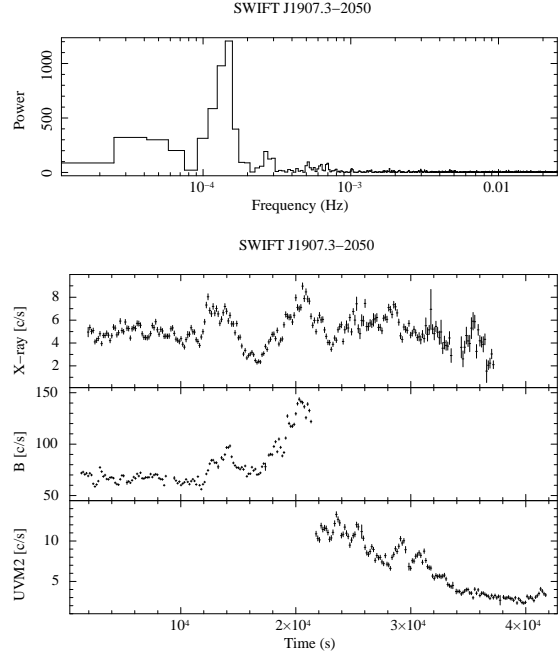

\begin{tabular}{c}
\includegraphics[angle=270,width=3.0in]{1907ps.ps} \\
\includegraphics[angle=270,width=3.0in]{1907lc.ps} \\
\end{tabular}
\caption{{\it Upper Panel}: \JA\ PN 0.3--15 keV power spectrum.
{\it Lower Panels:} \JA\ background subtracted light curve in 3 energy bands: PN 0.3--15 keV (top), B (center), UVM2 (bottom). The binning time is 150 s in all bands.}
\label{fig:1907lc}
\end{figure}


\section{Results}

\subsection{Swift J1907.3-2050}

\subsubsection{X-ray Timing Analysis}
\label{subsub:1907timing}

The PN 0.3--15\,keV light curve is highly variable (Fig. \ref{fig:1907lc}) with count rate 
 changing by more than $60\%$. A long-term  trend could  also be present, with a timescale larger 
than 8 h, but the exposure  (only 
 10.6 h) is not long enough to characterize this variability.
 The stronger variations are detected on shorter timescale ($\sim2$ h) after the first 10 ks since the start of 
the observation. 
The 0.3--15 keV power spectrum of the total observation (Fig. \ref{fig:1907lc}) shows an 
asymmetric peak at  $\sim$0.000139\,Hz with an excess at lower frequencies.
The first harmonic is also present and there is also indication of power at a lower frequency 
($\sim4\times10^{-5}$ Hz).  
The main peak would correspond to a period of $7200\pm500$ s ($2.00\pm0.14$ h).  All uncertainties are 
hereafter  at $1\sigma$ confidence level if not otherwise specified. 
 However, a closer inspection of the light curve in three time intervals: 0--10 ks, 10--25 ks, and 25--38 ks does not
reveal pulses above 3$\sigma$ significance, except in the 10--25 ks interval, where fractional variability  is $32.7\pm0.6\%$.
Therefore, the $\sim2$ h X-ray signal does not appear to be coherent. 
The extracted  light curves in  the four energy bands 0.3--1, 1--3, 3--5 and 5--15 keV 
in the 10--25 ks time interval reveal that this transient variability  is energy dependent, with 
fractional amplitude  decreasing from  $38\pm1\%$ in the  0.3--1\,keV band to $28\pm1\%$ in the hard  5--15\,keV band.

\begin{figure}
\centering
\includegraphics[width=3.0in]{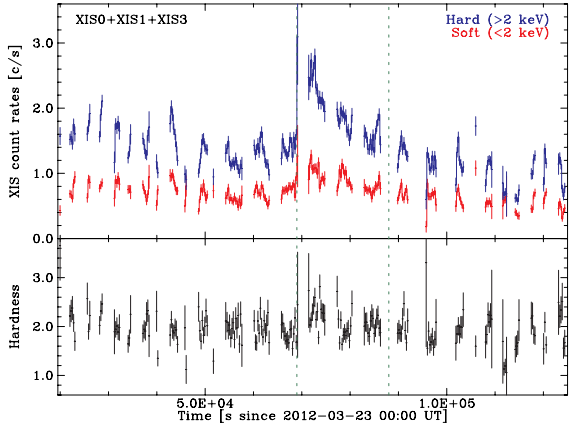}
\caption{ \JA\ : (Upper panel) \Suzaku/XIS light curves of \JA, in two energy bands.
(Lower panel) The hardness ratio as a function of time. The vertical dashed
lines define the flare interval used in Fourier analysis.}
\label{fig:1907suzakulc}
\end{figure}

\begin{figure}
\centering
\includegraphics[width=3.0in]{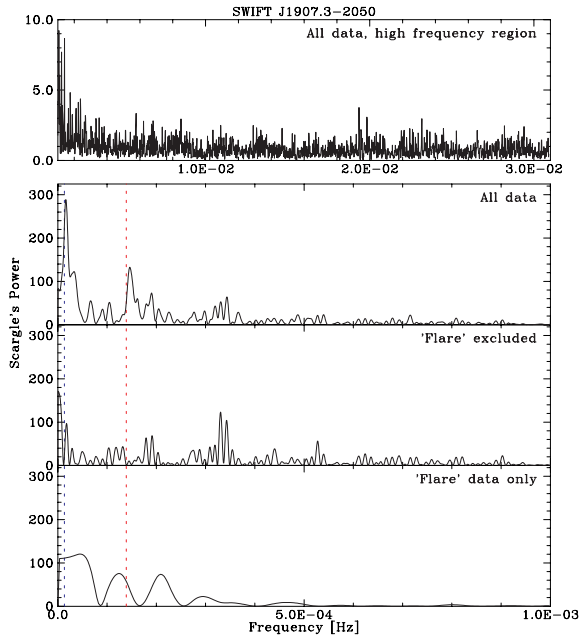}
\caption{ \JA\ : Power spectra of \Suzaku/XIS data using Scargle's definition
\citep{Scargle}.  The top panel shows the  high frequency region of the  power spectrum for all
data.  The bottom panels show the low frequency region, for the entire
data, the non-flare data, and the flare data only, from top to bottom. The vertical dashed lines correspond to 
the 2\,h variability detected in the \XMM\ data.}
\label{fig:1907suzakups}
\end{figure}

\noindent The \Suzaku\ X-ray light curve, spanning over $\sim$ 28.6 ks, includes a prominent brightening
(hereafter ``flare'') starting roughly halfway through the observation and
lasting for 3 spacecraft orbits (5--6 hours).  We have extracted the hard
($>$2 keV) and soft ($<$2 keV) light curves separately, but the hardness
ratio does not show a significant change during the flare
(Fig. \ref{fig:1907suzakulc}).  A Fourier analysis of the entire dataset
(without energy cut) shows substantial power at low frequencies, including
a peak near 2 hours, but no significant peaks at higher frequencies
(Fig. \ref{fig:1907suzakups}). We analysed the flare and non-flare data separately
and found that the 2 hour peak is not persistent through the observation;
it is definitely absent in the non-flare data.  While some power near this
frequency may well be present in the flare data,  no definite
conclusion can be made due to the relatively short duration of the flare and the \Suzaku\  orbital gaps.

\begin{figure}
\centering
\includegraphics[angle=270,width=3.3in]{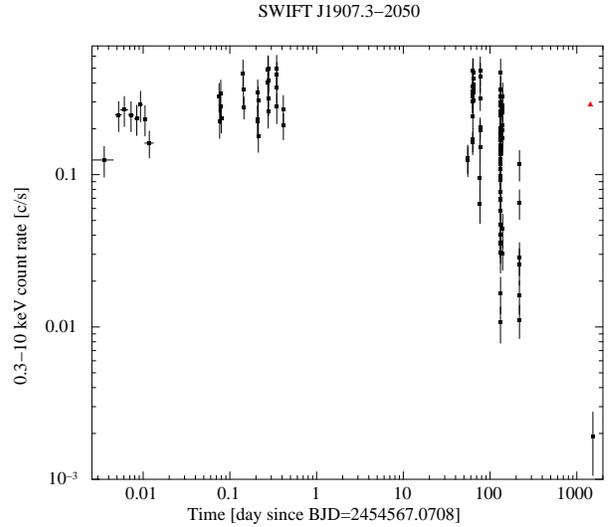}
\caption{\Swift/XRT light curve of \JA.   Due to the extremely sparse coverage, for plotting purposes, data are shown on a 
logarithmic scale.
Black squares represent \Swift\ data. The red triangle represent the estimated 0.5--10 keV \textit{Swift} count rate for the \XMM\ and the \Suzaku\ 2012 pointings. 
We estimated it  with \textit{webpimms} using a power law multiplied by local and galactic absorption as a spectral model.}
\label{fig:1907totallc}
\end{figure}

\noindent We also inspected the \Swift/XRT PC data of \JA\ to search for similar variabilities. 
The source count rate 
 changes along the total baseline (1533 d) from a  maximum of $\sim0.5$ c/s to a minimum of 0.002 c/s (see Fig. 
\ref{fig:1907totallc}). In particular, we note that the two 2012  
\Swift/XRT observations carried out in june, three months after the \XMM\ and \Suzaku\ pointings, show the lowest count 
rate (0.002 c/s). 
Furthermore, hints for short  term periodic-like variability, on the order of two hours, 
 consistent with that measured in the \XMM\ data, is found  in a  selected light-curve sub interval, where the \Swift\ 
 coverage was more dense (between d$\sim132$ and d$\sim132.4$  reference time  BJD=2454567.0708). However,
 due to the satellite gaps we cannot further constrain this variability.

\begin{table}
\centering
\caption{Summary of the main timing results for the three sources.  From left to right: P$^X_{1,2}$ (X-ray  periods); P$^X_{lt}$ (X-ray long term trend)}
\begin{tabular}{ccc}
\hline
Source        & P$^X_{1,2}$ $^a$  & P$^X_{lt}$ $^b$   \\
              & s                &   h            \\
\hline             
\JA          & $7200\pm500$ $^c$   & $>8.4$       \\
\JB\ &  -  & $7.6\pm0.1$ \\
\JC\ & $6100\pm120$,$4960\pm80$ &  $11.4\pm0.2$  \\
\hline 
\end{tabular}  
\label{tab:period}                   
\begin{flushleft}
$^a$ 1$\sigma$ cl.\\
$^b$ 3$\sigma$ cl. P$^X_{lt}$ must be interpreted as an indication of long term  variability.\\
$^c$ This signal is not coherent.
\end{flushleft}
\end{table}

\subsubsection{The UV and optical light curves}
\label{subsub:1907optical}

Swift\,J1907 is at 13.14$\pm$0.09 mag and at 14.70$\pm$0.04\,mag 
in the UVM2 and the B band, respectively, corresponding to fluxes of
$\rm 1.43 \times 10^{-14}\,erg\,cm^{-2}\,s^{-1}\,\AA^{-1}$ and to
$\rm 1.03 \times 10^{-14}\,erg\,cm^{-2}\,s^{-1}\,\AA^{-1}$.

\noindent The UV and optical light curves are  also highly variable (see Fig \ref{fig:1907lc})  with a 
strong rise  in the B band count rate after 12 ks, coincident in time with the X-ray flux increase.
 Both the X-ray and B light  curves display a first maximum at $\sim$13 ks and a second at $\sim$20 ks.
 In the subsequent UV exposure  the count rate decreases by a factor of 2.2 at the end of the pointing.     

\noindent We also divided the B band light curve in two intervals, I: 0--10 ks , and II: 10--22 ks, while 
we selected for the UVM2  light curve only the 22--38 ks interval (interval III).
Then, we cross-correlated the X-ray (used as reference) with the B and UWM2 band light 
curves in the three intervals. The cross-correlation function (CCF) in interval II is broad 
and asymmetric towards negative lags, with degree of correlation 0.6. For CCFs with complex shapes or asymmetries
the centroid is best evaluated using time lags in excess of 0.8\, the peak value \citep[e.g.][]{Peterson98,Zhang02}. 
We then find $\rm t_{cent}\simeq$ -310 s. The significance of the correlation is 15\,$\sigma$.
Instead, no correlation is found in interval I, while for interval II a weak degree of correlation is found (0.2).
Hence, we can  then conclude that the optical light leads the X-rays in the II interval only and that the transient signal
is also present at optical wavelengths.

\subsubsection{X-ray Spectral analysis}
\label{subsub:1907spec}

\begin{figure}
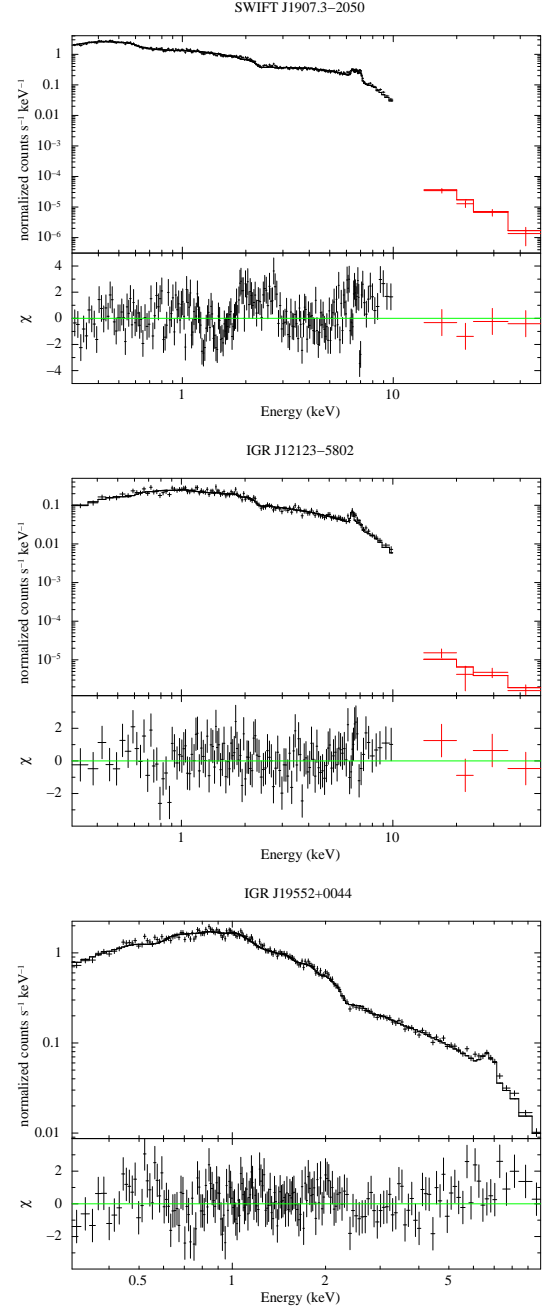

\begin{center}
\begin{tabular}{c}
\includegraphics[angle=-90,width=2.9in]{1907spec.ps} \\
\includegraphics[angle=-90,width=2.9in]{1212spec.ps} \\ 
\includegraphics[angle=-90,width=2.9in]{1955spec.ps} \\
\end{tabular}
\caption{{\it Upper panels:} \JA\ broad band 0.3-100 keV count spectrum. Data are rebinned for plotting 
purposes. XMM data are in black while \Swift/BAT data are in red. Residuals 
are shown in the bottom panel. {\it Center panels:} The same as above, but for \JB. Here the {\sc Cemekl} model is shown. {\it Lower panels:} The same as above, but for \JC, in the range 0.3-10 keV only (no BAT data are available). Here the {\sc Cemekl} model is shown.} 
\label{fig:spec}
\end{center}
\end{figure}

 The combined \XMM\ EPIC and \Swift\ BAT (0.3--100\,keV) average spectrum
was first fitted by a single optically thin component ({\sc mekal}) multiplied by a complex 
absorption that includes a total absorber ({\sc Wabs}) and a partial covering absorber ({\sc Pcfabs}).
 The latter is justified by the energy dependence of the transient signal. 
This gives a poor fit ($\chi^2_{\nu}$=2.05 for 1000 dof).
Substantial improvements are found  including additional components and verifying their statistical significance 
(greater than $3\sigma$)   with an F-Test. 
 We then obtain the best fitting model (see Fig. \ref{fig:spec}) consisting of two {\sc mekal} components, a cold (c) at 
0.12$\pm$0.02\,keV and a hot (h) at 13.2$\pm$0.6\,keV, plus  an optically thick ({\sc BB}) component at 61$\pm$5\,eV
and a Gaussian,  multiplied by a total and partial covering absorbers  ($\chi^2_{\nu}=1.15$ for 995 dof). No further component 
 is required by the fit. The spectral parameters are reported in  Tab. \ref{tab:1907avspec}. The black body component, 
significant at  $\>8\sigma$ confidence level,  has a radius of 40$^{+20}_{-10}$ km
(at 1 kpc)\footnote{A distance of $1.15^{+0.67}_{-0.42}$\,kpc has been estimated by \citet{Thorstensen2010}.}. 
The Gaussian accounts by for the prominent emission feature at 6.4\,keV with no significant energy shift. 
The total absorber column density is consistent with the  galactic value in the direction of the source derived from \cite{Kalberla05} and 
\cite{Dickey90}. No significant change with time is found 
in the spectral parameters within their statistical uncertainties.
 
\noindent The spectral analysis reveals that the X-ray emitting region is small and the plasma reaches typical temperatures 
achieved in a magnetically confined accretion flow, where a standing shock is formed at the poles of the compact star. 
Since the derived temperature of the hot component has to be  regarded as a lower limit to the maximum temperature of the 
post-shock region (PSR), we also used the  model of \cite{suleimanov05} (private code to be used into $Xspec$). 
This model takes into account the growth of pressure toward the WD surface and hence the change  of gravity, allowing 
us to obtain a more reliable  value for the maximum temperature and consequently an estimate of the WD mass. Since 
the  model is computed for the continuum only, we then added a broad Gaussian to take
into account the iron complex (thermal and fluorescence). A fit to the \XMM\ EPIC and \Swift\ BAT 
spectrum for $E>3$ keV gives a mass $\rm M=0.64^{+0.03}_{-0.04}\,M_{\odot}$ ($\chi^2_{\nu}=1.48$, 636 dof). 

\noindent We  inspected spectral parameters variability by comparing the \XMM\ EPIC and \Swift XRT spectra.  
Due to the low source count rate we accumulated an average XRT spectrum over 
all the \Swift\ pointings in 2008, excluding the 2012 ones where the count rate is much lower.
We used same model used for the EPIC  spectrum, but did not include the 6.4\,keV iron line and the {\sc BB} component
because they are not statistically required. We left free to vary all parameters except for the abundances that
were fixed a the \XMM\ best fit values. The spectral fit has a $\chi^{2}_{\nu}=1.08$ for 114 dof. 
The temperature of the hot {\sc mekal} is consistent, within uncertainties, with that found from the \XMM\  
 spectral fit, ($\rm kT_{h}=14_{-2}^{+4}$\,keV), while the temperature of the cold {\sc mekal} is 
unconstrained, likely due to the low S/N. A $3\sigma$ upper limit is $\rm kT_{c}<0.11$ keV.
 The 0.3--10 keV \Swift\ flux ($\sim7.9\pm0.3\times10^{-12}$ \ergscm) is a  factor  of $\sim3$ fainter  than 
that of \XMM  (see Tab. \ref{tab:1907avspec}). We also extracted the June, 2012 XRT spectrum but used a simple power-law, 
due to the extremely low S/N. At this epoch the flux is: $\rm 6.1^{+7.9}_{-3.3}\times10^{-14}$ \ergscm,
 indicating  a decrease by a factor of $\sim100$, in a three-months timescale. 
The \Suzaku\ spectrum when fitted with the same components of \XMM\ spectrum gives an average 0.3-10 keV flux of 
$2.3\times10^{-11}$ \ergscm, consistent with that obtained from the \XMM\ data.   From the long-term X-ray history 
(see also Fig. \ref{fig:1907totallc}), we then conclude that \JA\ is  also a highly variable X-ray 
source on months-years timescale.

\begin{table}
\caption{\JA: Spectral parameters of the best fitting model.
 Uncertainties are 1 sigma confidence level.}
\begin{center}
\begin{tabular}{ccc}
\hline 
\multicolumn{3}{c}{\JA} \\
\multicolumn{3}{c}{{\sc BB}+{\sc 2 mekal}}\\
\hline 
N$_{H_{\rm W}}$             &  $10^{22}$ cm$^{-2}$  &  $0.103\pm0.006$   \\ 
N$_{H_{\rm Pc}}$ 	        & $10^{22}$ cm$^{-2}$   & $9.6\pm0.5$        \\
cvf                         &  \%                  &  $67\pm1$          \\
kT$_{\rm BB}$               &  eV             & $61\pm5$    \\
kT$_{\rm c}$                &  keV                 & $0.12\pm0.02$      \\
kT$_{\rm h}$                & keV                  & $13.2\pm0.6$       \\
R$_{\rm BB}$                & km                   & $40^{+20}_{-10}$  \\
norm$_{\rm c}$              & $10^{-3}$            & $5^{+0.3}_{-0.9}$        \\
norm$_{\rm h}$              & $10^{-3}$            & $20.1\pm0.5$       \\
A$_{\rm Z}$                 &                      &  $0.70\pm0.05$     \\
EW$^{*}$                    & keV                  &  $0.15\pm0.01$     \\
F$_{0.3-10}$                &  $10^{-11}$          & $2.45\pm0.05$      \\
F$_{15-100}$                &  $10^{-11}$          & $1.3\pm0.1$        \\
F$^{BB}_{bol}$ $^{**}$      &  $10^{-11}$          &  $\sim2.9$        \\
F$^{c+h}_{bol}$ $^{***}$    &  $10^{-11}$          &  $\sim12$        \\
$\chi^2_{\nu}$ (dof)        &                      &  1.15 (995) \\
\hline 
\\
\end{tabular}  
\label{tab:1907avspec}                                                                                                         
\end{center}
$^{*}$ Gaussian energy fixed at 6.4 keV.\\
$^{**}$ Unabsorbed  bolometric flux  of the black body component.\\
$^{***}$ Unabsorbed bolometric flux  of the two optically thin components. \\
\end{table}

\subsubsection{The RGS spectrum of \JA}

\begin{figure}
\centering
\includegraphics[width=3.3in]{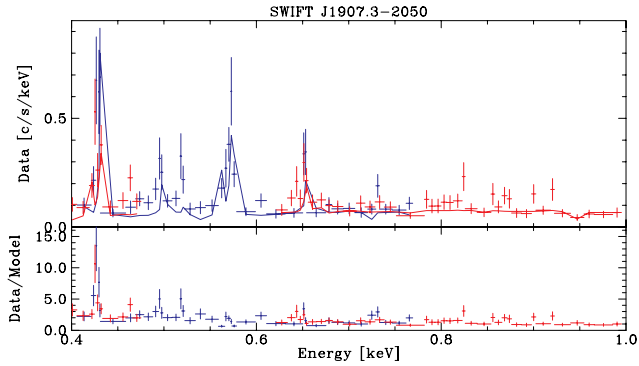}
\caption{\XMM/RGS spectra of \JA\ in the 0.4--1.0 keV region.  RGS1
data are shown in blue, RGS2 in red.  The top panel shows data with a
model in which nitrogen abundance was left as free parameter, with a
best-fit value of $\sim$2, while the abundances of other elements remain
fixed at 0.7.  The bottom panel shows the data to model ratio when the
abundance of all elements were set to 0.7.}
\label{fig:v1082rgs}
\end{figure}

We have applied the  best fit EPIC model (Tab. \ref{tab:1907avspec})
to the RGS spectra of \JA\ (Fig. \ref{fig:v1082rgs}). Although the overall trend
is reproduced by the model, there are significant residuals and $\chi^2_{\nu}$ 
remains relatively high ($\sim$1.3). The most prominent residuals
correspond to the He-like lines of nitrogen (E$\sim0.425$ keV).  Using the variable abundance
version of {\sc mekal} model, fixing the abundances at 0.7 except  that of
nitrogen,  we find a significant overabundance of
this element ($\sim$2 times Solar). While the $\chi^2_{\nu}$ is still not satisfactory ($\sim$1.2),
 the modest quality of the RGS spectra and the large number of free parameters 
do not allow further possible refinement.

\subsection{Swift\,J1212.3-5802}

\subsubsection{X-ray, UV and Optical timing analysis}

\begin{figure}
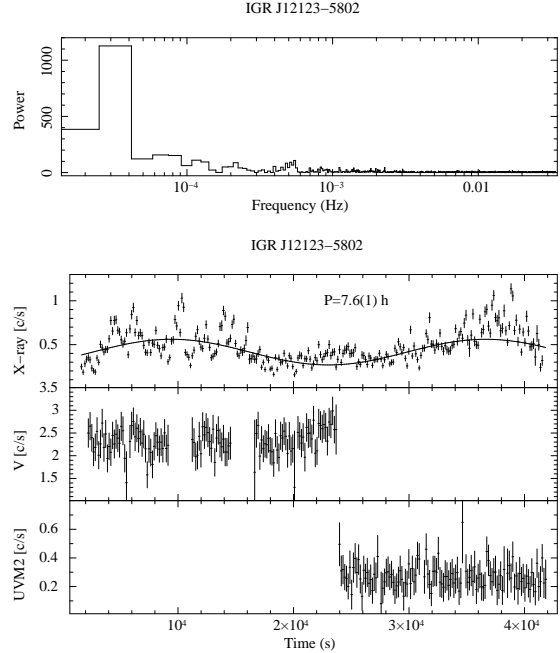

\begin{tabular}{c}
\includegraphics[angle=270,width=3.0in]{1212ps.ps} \\
\includegraphics[angle=270,width=3.0in]{1212lc.ps} \\
\end{tabular}
\caption{{\it Upper Panels}: \JB\ PN 0.3--15 keV power spectrum. 
{\it Lower Panels:} \JB\ PN plus MOS 0.3--15 keV background subtracted light curve (top), V band light curve (center), and UVM2 light curve (bottom). The binning time is 150 s in all bands. The solid line represents a fit made by a constant plus a sinusoid function (solid line).}
\label{fig:1212lc}
\end{figure}

The   combined EPIC PN and MOS light curve in the 0.3--15\,keV range (Fig. \ref{fig:1212lc}) shows the presence
 of a long term variability on which rapid changes by a factor of 2, occur on a timescale of tens of 
minutes. In the first 13 ks of observation the average count rate is $\rm \sim1.0\,cts\,s^{-1}$, 
between 13 ks and 30 ks the count rate is $\sim0.6$ c/s, while between 30 ks and 41 ks 
it is again $\sim1.0$ c/s. 
The   power spectrum  shows a peak at $\sim3.5\times10^{-5}$ Hz, $\sim8$ h (see Fig. \ref{fig:1212lc}).  
 A sinudoidal fit to the 0.3--15 keV light curve  gives a period of $7.6\pm0.1$ h, $3\sigma$ cl. 
 Given that the \XMM\ observation is 11.1\,h long, we cannot assess if this variability is truly periodic.

 \noindent  The power spectrum  of the light curve detrended from the main variability (7.6\,h) does not reveal 
significant peak.  A short term  variability seems to be present during the high rate sections of the  observation ($\rm 
t\lesssim18$ ks, $\rm t\gtrsim30$ ks)
 and the corresponding power spectrum peaks at 1900 s.  However, this peak is only present in the first 
section. Hence, it is likely that at  high rates \JB\  is affected by QPO-like variability. 
We also inspected the behaviour of HR with time, using  
both 1--3 keV/0.3--1 keV and 5--15 keV/3--5 keV ratios. No change is found within statistical 
uncertainties.

\noindent The OM V and UVM2 band photometry has low statistics, due to the faintness of 
the source (V=17.1\,mag and UVM2=17.2\,mag), and does not show the changes observed in the 
X-rays (Fig. \ref{fig:1212lc}). We also analysed ground-based B band differential 
photometry acquired on 24-26\, May 2011 at the 2.15\,m telescope at the Complejo Astronomico el 
Leoncito, \emph{CASLEO}, in  Argentina, equipped with a  direct CCD camera. \JB\ was observed for 
about 5\,h each night. Exposure times of individual images were 60, 90 and 120 s respectively. 
The B band light curve (not shown) displays flickering-type variability on timescales from few minutes to tens of 
minutes with no clear periodic trend during the three nights.

\subsubsection{X-ray Spectral analysis}

We found two spectral models that are equally statistically acceptable. 
The first one is composed by a multi-temperature optically thin plasma in which the emission measure follows 
a power-law in temperature ({\sc Cemekl}) plus a Gaussian at 6.4 keV to 
account for the fluorescent 6.4\,keV Fe line. The power-law  index  $\alpha$,  when left free to 
vary is  consistent within errors to unity. We, therefore, fixed it at this value.
The second one is made by the sum of two {\sc mekal} and a Gaussian at 6.4 keV. Both models 
required a complex absorption  made of  a total ({\sc Wabs}) and  a 
partial covering ({\sc Pcfabs}) absorbers. In both cases, the inclusion of a black body component
is not statistically required, its significance being $ < 2\sigma$.
Spectral fit results are reported in Tab. \ref{tab:1212avspec} and shown in Fig. \ref{fig:spec}
({\sc Cemekl} model only).

\begin{table*}
\caption{\JB: spectral parameters of the best fitting models.
Uncertainties are 1 sigma confidence level. }
\begin{center}
\begin{tabular}{cccccc}
\hline 
\multicolumn{6}{c}{\JB} \\
\multicolumn{3}{c}{\sc Cemekl} & \multicolumn{3}{c}{\sc 2 mekal} \\
\hline 
N$_{H_{\rm W}}$       &  $10^{22}$ cm$^{-2}$ & $0.11\pm0.01$  & N$_{H_{\rm W}}$            & $10^{22}$ cm$^{-2}$ &  $0.11\pm0.01$ \\
N$_{H_{\rm Pc}}$      & $10^{22}$ cm$^{-2}$  & $2.3\pm0.4$    & N$_{H_{\rm Pc}}$           & $10^{22}$ cm$^{-2}$ & $2.8\pm0.5$    \\
cvf                   &  \%                 & $53\pm3$       & cvf                         &  \%                 &  $49\pm2$      \\
                      &                     &                & kT$_{\rm c}$                &  keV                & $5.95\pm0.02$  \\
kT$_{\rm max}$        &   keV               &  $>$43         & kT$_{\rm h}$                &  keV                & $>$62         \\
norm                  & $10^{-3}$           & $6.4\pm0.3$    & norm$_{\rm c}$              & $10^{-3}$           & $0.5\pm0.2$    \\
                      &                     &                & norm$_{\rm h}$              & $10^{-3}$           & $2.6\pm0.5$    \\
A$_{\rm Z}$           &                     & $1.0\pm0.3$    & A$_{\rm Z}$                 &                     &  $1.5\pm0.4$   \\
EW$^{*}$              & keV                 & $0.25\pm0.03$  & EW                          & keV                 &  $0.21\pm0.03$ \\
F$_{0.3-10}$          &  $10^{-11}$         & $0.46\pm0.01$  & F$_{0.3-10}$                &  $10^{-11}$         & $0.46\pm0.02$  \\
                      &                     &                & F$_{15-100}$                &  $10^{-11}$         & $0.77\pm0.01$  \\
F$_{bol}$ $^{**}$  &  $10^{-11}$         &  $1.5\pm0.1$   & F$^{c+h}_{bol}$ $^{***}$ &  $10^{-11}$         & $1.6\pm0.1$     \\
$\chi^2_{\nu}$ (dof)   &                     &   0.93 (261)     & $\chi^2_{\nu}$ (dof)   &      &  0.87 (259)       \\
\hline 
\\
\end{tabular}   
\label{tab:1212avspec}
\end{center}
\begin{flushleft}
$^{*}$ Energy of the Gaussian fixed at 6.4 keV.\\
$^{**}$ Unabsorbed  bolometric flux  of {\sc Cemekl} model. \\
$^{***}$ Unabsorbed bolometric flux of the {\sc mekal} components. \\
\end{flushleft}
\end{table*}

\noindent The hydrogen column density of the total absorber is lower than the total 
galactic column  density in the direction of the source (3.2--4.1 $10^{22}$ cm$^{-2}$),
 suggesting an interstellar origin. On the other hand the $\rm N_H$ of the
partial covering absorber (cvf$=49-53\%$) indicates it is local.
For both models, {\sc Cemekl} and two-{\sc mekal}, the  high temperature component  is unconstrained. 
Lower limits of  43 keV and 62 keV are found, respectively. The 6.4\,keV fluorescent Fe line 
is strong in this source (EW$=250$ eV) indicating that reflection is not negligible. 
However, a reflection component is not statistically significant in the fits. 
In both cases  the abundances are, within uncertainties,  consistent with  the solar value. 
We are unable to prefer one of the two models with current data.
To estimate the WD mass, we applied  also for this source the \citet{suleimanov05} model obtaining 
$\rm M_{WD}=0.89^{+0.02}_{-0.04} M_{\odot}$.
The 0.2--10 keV spectrum, extracted between 0--14\,ks (high rate) and 14--30\,ks (low rate), 
if fitted with the same model does not reveal spectral changes within statistical uncertainties, 
except for the  normalization. 
Hence, the large variability observed during the \XMM\ exposure is due to changes of the emitting 
volume.

\subsection{IGR J19552+0044}

\subsubsection{X-ray timing analysis}

\begin{figure}
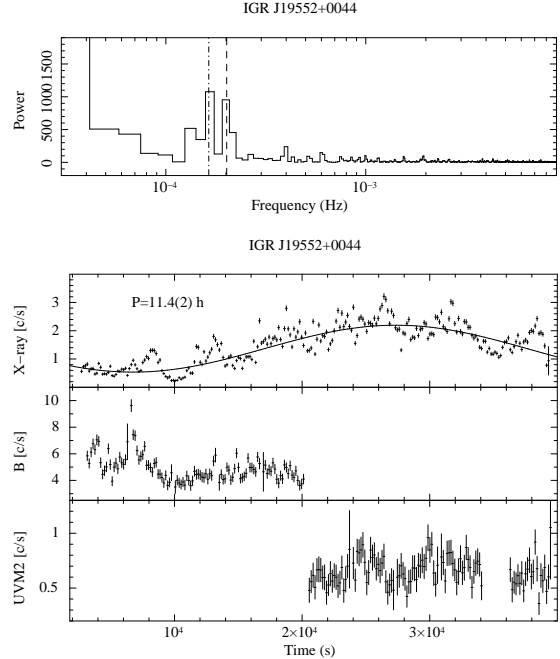

\begin{tabular}{c}
\includegraphics[angle=270,width=3.0in]{1955ps2.ps} \\
\includegraphics[angle=270,width=3.0in]{1955lc.ps} \\
\end{tabular}
\caption{{\it Upper Panel}: \JC\ PN 0.3--15 keV power spectrum. The dashed and dot dashed 
 lines at $\sim1.7\times10^{-4}$ Hz and $2\times10^{-4}$ represent the 1.69 h and 1.32 h periods,
respectively. 
{\it Lower Panels:} \JC\ PN plus MOS 0.3--15 keV background subtracted light curve (top), B band light  curve (center), and 
UVM2 light curve (bottom). The binning time is 150 s in all bands. The solid line represents a fit made  by a constant plus a 
sinusoid function (solid line).}
\label{fig:1955lc}
\end{figure}

IGR\,J1955 is also a highly variable source. The 0.3--15 keV combined EPIC PN and MOS light curve  displays a large amplitude 
variation over the whole observation where count rate increases by a factor $\sim$3 (Fig. \ref{fig:1955lc}).
A  dip lasting $\sim1.5$ ks is also observed at $\sim$10 ks since the start of the observation.  The
 count rate decreases by $\sim$9.7 times the average value, almost reaching zero counts.
A short term (few hours) variability is also present.

\noindent  The  X-ray  power spectrum shows three main peaks: a short frequency one at $\sim2.5\times10^{-5}$ Hz ($\sim11$ h), and two close ones 
at $\sim1.7\times10^{-4}$ Hz and $2\times10^{-4}$ Hz ($\sim 6000$ s and $\sim5000$ s) respectively (see Fig. \ref{fig:1955lc}).
 A sinusoidal fit to the light curve gives a period of $11.4\pm0.2$ h ($3\sigma$ cl). 
This value should be considered as an indication of  a long term variability, because of the limited  \XMM\ exposure.
For the two short term variabilities, both  significant at $8\sigma$ cl, we obtain: $6100\pm120$ ($1.69\pm0.04$ h) 
and $4960\pm80$ s ($1.32\pm0.08$ h). On the other hand the  UV and B band light curves are too short to allow a timing 
study. However, we can identify an optical counterpart (B band), although much shallower,  to the dip  observed in the 
X-rays (see Fig. \ref{fig:1955lc}).

\begin{figure*}
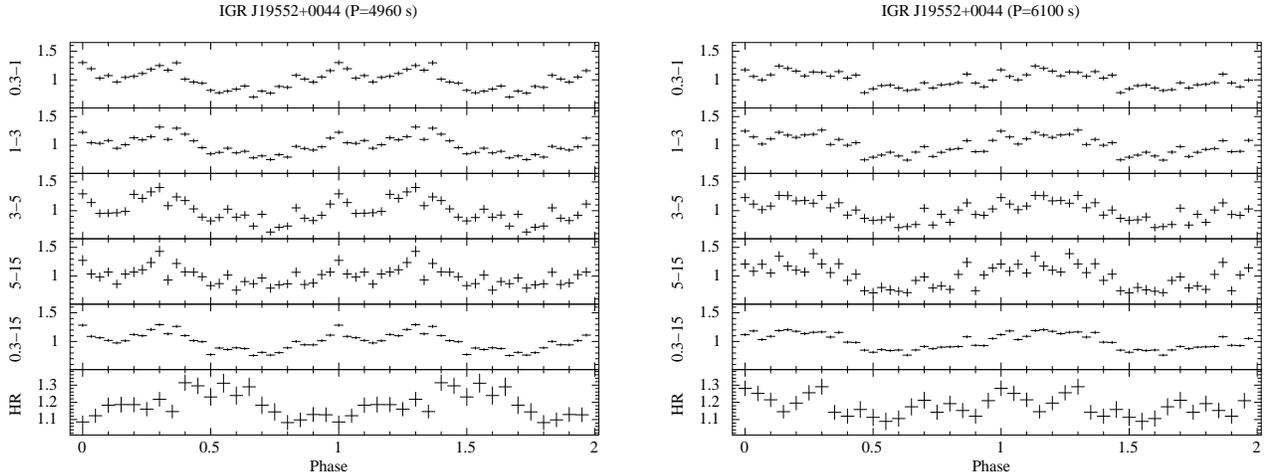

\centering
\begin{tabular}{cc}
\includegraphics[angle=270,width=3.3in]{1955_pf_complete_4960.ps} &
\includegraphics[angle=270,width=3.3in]{1955_pf_complete_6100.ps} 
\end{tabular}
 \caption{{\it Left panels:} From top to bottom, \JC\ pulse profiles folded at 4960 s in five energy intervals, 0.3--1, 1--3, 3--5, 5--15, 0.3-15 keV, on the y axis the normalized intensities. On the bottom line the Hardness Ratio (HR) for the 1--3 keV vs 0.3--1 keV band. {\it Right panels:} The same as left panels, but for the 6100 s period.}
 \label{fig:1955pf}
\end{figure*}

\noindent We studied the energy dependence of the pulse shape, by fitting the  
light curves in the ranges 0.3--1 keV, 1--3 keV, 3--5 keV and 5--15 keV at the two periods.
The pulse at 4960 s is phase aligned and structured at all 
energies, with a double 
peaked maximum at phases $\sim0.0$ and $\sim0.3$ respectively. The pulse fraction PF\footnote{The pulse fraction is here  
defined as: PF=$\rm (A_{max}-A_{min})/(A_{max}+A_{min})$ where $\rm A_{max}$ and $\rm A_{min}$  are respectively the  maximum and minimum value of a 
sinusoidal function.}
 slightly decreases with energy: PF$_{0.3-1}=18.2\pm0.8\%$, 
PF$_{1-3}=16.5\pm0.8\%$, PF$_{3-5}=20\pm2\%$, PF$_{5-15}=13\pm2\%$. The average PF in the 0.3--15\,keV  range is 
$17.0\pm0.5\%$. The hardness ratio (HR) between 1--3\,keV and 0.3--1\,keV  indicates a tendency of 
spectral hardening (significant at $3.6\sigma$ 
level) close to  pulse minimum,  $\phi$=0.4--0.6  (see Fig. \ref{fig:1955pf}).  
The low S/N does not allow to infer changes in the HR between 5--15 keV and 3--5 keV bands.

\noindent The pulse at 6100\,s is instead less structured and roughly 
 single peaked, with the maximum at $\phi=0.2$, phase aligned at all energies.
 The PF slightly increases with energy: PF$_{0.3-1}=14.6\pm0.8\%$, 
PF$_{1-3}=18.5\pm0.8\%$, PF$_{3-5}=20\pm2\%$, 
 PF$_{5-15}=22\pm2\%$. The average PF in the total 0.3--15 keV band is $17.1\pm0.5\%$. 
A marginally significant ($2.4\sigma$) spectral hardening at pulse maximum is found 
from the HR between 1--3 keV and 0.3--1 keV bands.

\noindent Finally,  the HR between 1--3\,keV and 0.3--1\,keV over the whole \XMM\ observation reveals a hardening 
when flux increases.  The ratio between the two light curves is  $0.78\pm0.03$ during the first 6 ks, while later 
 it is $1.18\pm0.02$.

\subsubsection{X-ray spectral analysis}

 The best  fit model is made of a multi-temperature ({\sc Cemekl}) plasma absorbed by a total absorber 
({\sc Wabs}) and a Gaussian to account for the excess at 6.4 keV.
 The inclusion of a partial covering absorber ({\sc Pcfabs})  is not statistically required. 
The power law $\alpha$ index is $0.74\pm0.05$ and its inclusion as free parameter is significant at $4.8\sigma$.
 The N$_H$ column density is $0.07\pm0.02\times10^{22}$ cm$^{-2}$, and is lower than the total galactic 
column density in the direction of the source  ($0.13-0.16\times10^{22}$ cm$^{-2}$). 
 The fluorescent 6.4\,keV Fe line is moderately weak (EW= 70$\pm$20\,eV).
A soft black body component is not required.
All model parameters  are reported in Tab. \ref{tab:1955avspec} (see also Fig. \ref{fig:spec}). Similarly to the other 
sources, we also  applied the \cite{suleimanov05} model to derive the WD mass, and find $\rm 
M_{WD}=0.77^{+0.02}_{-0.03}\,M_{\odot}$ ($\chi^2_{\nu}=1.09$, 489 dof).

\begin{table}
\caption{\JC: spectral parameters of the best fitting model. Uncertainties are 1 sigma confidence level}.

\begin{center}
\begin{tabular}{ccc}
\hline 
\multicolumn{3}{c}{\JC} \\
\multicolumn{3}{c}{{\sc Cemekl}}\\
\hline
N$_{H}$           &  $10^{22}$cm$^{-2}$  &  $0.076\pm0.003$  \\ 
kT$_{\rm max}$    &  keV                 & $36\pm4$          \\
norm              & $10^{-3}$            & $9.9\pm0.4$       \\
$\alpha$          &                      & $0.74\pm0.05$     \\
A$_{\rm Z}$       &                      &  $0.62\pm0.07$    \\
EW$^{*}$          & keV                  & $0.07\pm0.02$     \\
F$_{0.3-10}$      &  $10^{-11}$          & $1.06\pm0.01$     \\
F$_{bol}$ $^{**}$ & $10^{-11}$          &  $2.2\pm0.1$                  \\ 
$\chi^2_{\nu}$ (dof)     &          &  1.05 (725)   \\
\hline 
\\
\end{tabular}  
\label{tab:1955avspec}                                                                               
\end{center}
\begin{flushleft}
$^*$ Energy of the Gaussian fixed at 6.4 keV.\\
$^{**}$ Unabsorbed bolometric flux of {\sc Cemekl} model.\\
\end{flushleft}
\end{table}

\noindent  We also investigated phase-dependent spectral variability at the 4960\,s period, by selecting the phase intervals 
 $\phi=0.0-0.5$ and $\phi=0.5-0.9$ corresponding to the maximum and minimum respectively. 
  We fixed at their average  values the hydrogen column density N$_H$, the metal abundance $\rm A_Z$, and the Gaussian energy, 
 all other parameters were left free to vary. Only the normalization was found to change ($\sim20\%$)  according to  the flux 
 variation.

\noindent The RGS spectra have low S/N but broadly confirm the EPIC spectral fits.  
The strongest lines seen are those of OVII, OVIII, and FeXVII (0.83 keV=15A).
Some lines (N VII, O VII, Mg XI and Mg XII) may be stronger than that
in the best fit  model. However, this should be re-examined with higher S/N data.

\section{Discussion}

\subsection{Swift\,J1907.3-2050: a long period magnetic CV?}

The X-ray history of \JA\  over $\sim$5 yr reveals a highly variable X-ray source on both 
short (hours) and long (months-years) time scales. 
We detect a  transient feature with characteristic timescale of $\sim$2 h in the \XMM\ 
X-ray and optical data. It may also be  present in the \Suzaku\ and \Swift\ XRT data 
at some epochs. The possibility that it represents the spin period of the WD is highly unlikely, 
since the spin signal is not expected to disappear without a significant change in the source state.
The characteristic 2 h timescale is close 
to the free-fall time from  the inner Lagrangian point. Indeed, with a WD mass of $0.64\rm M_{\odot}$ and a 
donor star mass of $0.8\rm M_{\odot}$ \citep{Thorstensen2010},  this binary would have an inverted mass 
ratio $\rm q = M_{sec}/M_{WD}=1.3$. The WD Roche Lobe radius would be $1.8\times 10^{11}$ cm and the 
free-fall time $\sim$5900 s. The occurrence of this feature at high rates  (both in the X-rays and UV/optical) 
suggests a transient enhancement of mass transfer from the donor star overflowing the accretion disc.
For the latter we estimate a color temperature of 12\,kK from the lowest observed UV and B band 
fluxes corrected for reddening 
\footnote{ We adopt an extinction of $\rm E_{B-V}=0.14$, as derived 
from the hydrogen  column density from X-ray spectral fits \citep{Ryter75}.}.
The flare activity also indicates that mass accretion rate changes are not uncommon on timescales of hours.

\noindent \JA\ also varies on a longer time scale ($\gtrsim$8 h) in both X-ray and optical/UV ranges 
and we could detect a flare-like activity lasting $\sim$6\,h in the \Suzaku\ data.  With the present data we
cannot assess whether it is related to the long 20.82 h orbital period.
 Due to the lack of energy dependence the flare is likely due to a temporary enhancement in the accretion rate, rather 
than resulting from an orbital modulation. 

 \noindent \JA\ was already  known to display high, intermediate and low optical states, 
where accretion partially or completely switches-off in some occasions \citep[see][and reference therein]{Thorstensen2010}.
The X-ray history  also shows that it undergoes high and low X-ray states on a months-years time scale, thus
corroborating mass accretion rate changes. This peculiarity is typical of 
 magnetic systems of the Polar type or nova-like CVs of the VY\,Scl type stars \citep[see e.g.][]{honey04}.  
  The orbital period of 20.82 h is by far longer than those of VY\,Scl stars, which cluster at the 3 h edge of the orbital period gap, and also longer 
than those of  Polars, being V1309\,Ori (8 h) the longest period Polar known so far.

\noindent  The spectral characteristics of \JA\ may favour a magnetic system  as early suggested by 
\citet{Thorstensen2010}. A high complex 
absorbing material, is a defining characteristic of magnetic systems, particularly of 
IPs,  rather  than of  non-magnetic CVs \citep[see][]{Ramsay2008}.  Also a  
multi-temperature plasma  reaching high temperatures ($\sim14$ keV) is typical 
of accretion  shocks at the WD poles.  An optically   thick component, with temperature of 60 eV and with a small emitting 
area,  is also a characteristic of magnetic systems only.  This component  is a recognized feature  in Polars and recently 
in IPs.  However, an orbitally-locked WD rotation at the extremely 
long 20.82 h period would imply a very high magnetic field, $\gtrsim1.4\times10^{10}$ G  
\citep[see][]{Warner95}. 
No WD in CVs has ever been detected with such magnetic field strength, being the highest field Polar AR\,UMa with B$=2.3\times10^{8}$ G.  
At high field strengths accretion shocks are cyclotron dominated  and even buried at low local 
accretion rates \citep{woelk_beuermann96,fischer_beuermann01}. This would yield 
a strong soft component. We instead infer a bolometric flux ratio 
between the black body and optically thin components $\rm F_{BB}/F_{thin}\sim0.24$.\\  
The remaining possibility is that \JA\ is an IP.  This would be consistent 
with  the spectral characteristics of this sub-type of magnetic systems.   Known IPs at the 
longest ($\gtrsim10$ h)   periods,  such as GK\,Per (47.9 h),  RX\, J1730-0559 (15.4 h) and 
AE\,Aqr (9.9 h), possess  rapidly spinning 
 WDs with  high degree of asynchronism ($\leq$0.002), with the only  exception of  V1062\,Tau 
(10 h, and $\sim0.1$).  However, no rapid X-ray pulses are detected in the X-ray power spectrum of \JA. 

 \noindent  The presence of a black body component at 60\,eV, with a ratio between the bolometric 
soft and hard X-ray fluxes of about 0.24, is a key signature of magnetic accretion. 
 Typical range of temperatures  in Polars is 20--40 eV, with a few systems at $\sim$60\,eV. 
In IPs the soft component is instead found to span a wider range of temperatures up to $\sim$90 eV  
 \citep{Anzolin08,Anzolin09,Bernardini12}. 
Hence \JA\ is  located  in the soft-to-hard X-ray luminosity ratio vs temperature plane 
\citep[see][]{Anzolin08} midway between the Polars and IPs.
In \JA\ this component  arises from a small region of 40 km 
(a 3$\sigma$ upper  limit is 114 km at 1 kpc).  
For a WD mass of 0.64 $\rm M_{\odot}$  the corresponding radius is 8.3$\times 10^{8}$\,cm  
\citep{Nauenberg72},  implying that the reprocessed   optically  thick emission comes from a 
fractional area $f\sim6\times10^{-6}$ (3$\sigma$ upper limit is $f\lesssim2\times10^{-4}$). 
It was recently demonstrated that this soft X-ray component is constituted by a range 
of temperatures over the accretion spot area \citep{Beuermann08,Beuermann12}. The projected fractional
WD surface area in the prototypical Polar AM\,Her is about 0.07$\%$  with the hottest regions ($>$ 40\,eV)
arising from much smaller fractional areas, $< 2\times 10^{-6}$ \citep[see][for detail]{Beuermann12}.  
 For \JA\ we can only infer the average temperature of this component and
a rather small and weakly constrained emitting area.  These values are closer to those 
derived in the soft IPs  \citep{Anzolin08}, which seem to possess smaller and hotter spots than Polars. 
Hence,  though sharing  
similar properties  of soft emitting magnetic CVs, it cannot be  unambiguously ascribed to either the IP or  
the Polar groups.

\noindent  To help in understanding this CV, we also estimated its mass accretion rate. 
We assume that the accretion luminosity is represented by the X-ray bolometric luminosity,
$\rm  L_{acc} = G\,\dot M \,M_{WD}\,R_{WD}^{-1} \sim L_{BB} + L_{thin}$, that includes 
both reprocessed  X-ray component and that of  the hard X-rays.
Using the derived WD mass and radius values of 0.64 $\rm M_{\odot}$ and $8.3\times10^{8}$ cm,   
we estimate $\rm \dot{M}=2-4\times10^{-9}\, M_{\odot}\,yr^{-1}$, 
for a minimum distance of 730 pc and a maximum of 1.15 kpc respectively. 
This value is higher than the mass trasfer rate ($\rm 
\dot{M}_{sec}\sim5\times10^{-10}\,  M_{\odot}\,yr^{-1}$)  
expected for a donor filling its Roche lobe \citep{Warner95} at a 18-20\,h orbital period. It is however lower 
by two orders of magnitude than that predicted by magnetic braking \citep{McDermott&Taam} at the same orbital periods. Consequently, it may be
possible that \JA\ has recently entered into a CV phase. 
The high  nitrogen abundance is also a peculiar feature of \JA\ . A number of CVs were found from
 far-UV spectra to display anomalous CNO abundance and believed to be descendant of massive progenitor donors 
\citep{Gaensicke03}.
This, together with  the large binary mass ratio and the long orbital period, 
might suggest that \JA\ is also descending from a  progenitor binary which underwent a phase of thermal time-scale mass transfer  
\citep[TTMT][]{schenker04,Podsiadlowski03}.

\noindent Further long observations in the X-rays and in the optical, including polarimetry as well as UV spectroscopy, will 
help to understand the true nature and evolutionary status of this CV.

\subsection{IGR\,J12123-5802 a true IP or a magnetic impostor?}

IGR\,J1212 is  a faint, but highly variable X-ray 
source. We find  a $\gtrsim$7.6\,h variability although we cannot assess whether it is is related to 
the binary orbit.  Variations  on time scales of 
$\sim$1900 s are not periodic. None of these variabilities have an optical counterpart where only flickering 
on timescale of a few minutes is found.

\noindent On the other hand, the broad-band X-ray spectrum is hard and well described by a multi-temperature 
plasma  with a high, although unconstrained, maximum temperature. We also infer a high mass for the WD 
($0.89\rm  M_{\odot}$). The spectrum is highly absorbed by interstellar and local material, the latter 
covering $\sim50\%$ the X-ray source. A strong (EW$=250$ eV) Fe fluorescent line is also present suggesting 
that reflection from cool material is present in this system. 
A high column density partial absorber and an intense Fe 
fluorescent line are defining characteristics of  magnetic CVs \citep{EzukaIshida99}. Hence, a magnetic system might be favoured. The lack of detection of a black body component also does not favour a Polar-type CV. 
While the lack of detection of a coherent periodicity does not allow us to 
classify \JB\ as an IP, we here note that other systems, such as LS\,Peg, EI\,UMa and 
V426\,Oph \citep{Ramsay2008}, have also failed to show coherent periodicities and 
 due to the spectral characteristics similar to IPs, 
were  defined as "bona fide" IPs. For  V426\,Oph and  LS\,Peg it is also speculated that 
the lack of pulsations could be due to the close alignment of the magnetic and rotation axes.

 \noindent The high temperature derived from X-rays if due to magnetic accretion would imply a mass 
of 0.89$M_{\odot}$, that would locate \JB\ within the range of most CV primary masses 
\citep{zorotovic11}. 
On the other hand if this temperature is achieved through a Keplerian disc accretion it would give a 
WD mass of $\sim$1.2$M_{\odot}$. Few CVs are known to harbour such massive WDs \citep{zorotovic11}.

\noindent Optical spectroscopy and polarimetry, as well as long photometric runs, could help to shed light into this new CV.

\subsection{IGR\,J19552+0044: a long period IP or a magnetic CV below the gap?}

 We have detected periodic variabilities at 1.69 h (6100 s), 1.38 h (4960 s). 
 A large amplitude variability on timescale of 11 h is also present. 
Support to both the short and the long term variabilities also comes from optical ground-based photometry and 
spectroscopy  (Tovmasian et al. in preparation, private communication).
  The X-rays arise from a multi-temperature plasma with maximum temperature of 36\,keV suffering little absorption
of interstellar  origin. The 1.38\,h periodicity is  due to changes in the normalization of the
emitting plasma.  These features are rather similar to HT\,Cam 
 \citep{deMartino05}, a low accretion rate IP below the orbital period gap.  
The bolometric X-ray luminosity is $\rm 4.3\times10^{31}\,d_{100pc}^{2}\,erg\,s^{-1}$. 
 If it represents the accretion  luminosity, for $\rm M_{WD} =0.8\,M_{\odot}$ and 
$\rm R_{WD}=7\times10^{8}$ cm \citep{Nauenberg72}, the accretion rate would indeed be very low:
$\rm 4.5\times10^{-12}\,d_{100pc}^{2}\,M_{\odot}\,yr^{-1}$. 

\noindent  The X-ray periodicities of 1.69 h, 1.38 h and the $\sim$11 h variability are, however, difficult to interpret. 
We here propose and discuss two possibilities. 

\subsubsection{A long period IP or a pre-Polar?}

If 1.38 h represents the spin of the accreting WD and 1.69 h
is a sideband period ($\rm P_{\omega - \Omega}$ or $\rm P_{\omega -2\Omega}$) , the orbital period would 
be either 7.3\,h or 14.6\,h. The detected variability at 11 h or longer, if periodic, would suggest 
the second case. However, such long-period system is expected to have a high accretion 
rate and a main sequence G3-5 type donor star \citep{Smith_Dhillon98}. A re-analysis of the 
 optical spectra by \citet{Masetti10} does not reveal G-type star features but only weak absorptions in the
red portion,  possibly suggesting Ca\,I 6162 \AA, and TiO 6900 \AA. Hence a G-type star is rather unlikely.
The 2MASS measures give only an upper limit to the J band magnitude (J$<$ 17.03), implying a 
 lower limit to the distance of 6.4\,kpc which seems to be in contradiction with the inferred low 
absorption. 
Furthermore, the ratios of He\,II and  H$_{\beta}$  line fluxes and equivalent widths (EWs)  
are closer to those found in Polars rather than in IPs 
\citep{vanparadis84} and the overall spectrum very much resembles those of Polars. 
Also, \JC\ is found at B=17.6\,mag, at similar flux level when observed by \citet{Masetti10} in 
2009, but much fainter than in the USNO (15.6 mag) and in the  Sloan Digital Sky Survey (SDSS) 
catalogues (g=15.9\,mag). 
A 1.6--2 mag difference suggests a highly variable source,  that is more common to short period 
CVs.

\noindent Alternative possibility is that the 1.38\,h and the 1.69\,h periods are the spin and the 
orbital periods, respectively.  A short period  magnetic CV could reconcile  the low 
accretion rate and  the absence of  intrinsic absorption  derived from X-rays. The beat period would then 
be 7.29\,h, not detected but the 
long term variability could be a sign of a sub-harmonic, although difficult to interpret. Clearly a longer 
observation would solve the true nature of the long-term variability. If it were the case the spin-to-orbital 
 period ratio would be 0.82. The period of 1.69 h would locate this source below the orbital period gap, where most Polars 
are found and only five IPs.
The latter have spin-to-orbit period ratio ranging from 
0.1 (HT\,Cam) to 0.68 (EX\,Hya). On the other hand, only five Polars are known to possess desynchronized 
primaries (CD\,Ind,  BY\,Cam, V1500 Cyg, V4633\,Sgr and V1432\,Aql) with spin-to-orbital period ratios 
between 0.98 and  1.02. \JC\ would then be either an extremely low asynchronous IP 
or the  most asynchronous Polar.  Only one system is known to share 
 a similar property, Paloma also know as RX\,J0524+42 \citep{Schwarz07}. Its spin-to-orbit period ratio of 
0.93--0.97  (depending on the true spin period) and its orbital period of 2.6 h locate it in the period gap. 
Paloma  is believed to be a transition object between IPs and Polars. 
As no IP is theoretically expected below the orbital period gap, the few IPs will likely
never synchronize \citep{norton04,norton08}. However, \JC\ being so 
close to  the line of synchronization could still evolve towards synchronism. In this case magnetic 
field measurements through polarimetry are crucial.

\noindent The X-ray analysis also reveals a hard X-ray spectrum but no soft X-ray black body component.   
Three of the asynchronous Polars mentioned above (CD\,Ind, 
BY\,Cam, and V1432\,Aql) display unusually hard  X-ray spectra and are detected above 20\,keV in the  
INTEGRAL/IBIS or  Swift/BAT surveys.  Interestingly also Paloma has been detected by 
\emph{Swift}/BAT and has been recently catalogued in the latest 70-month BAT catalogue 
\citep{Baumgartner13}.

\subsubsection{A magnetic evidence from IR}

IGR\,J1955 is reported in the 2MASS catalogue with J$<$17.03 mag, H$=15.63\pm0.11$ mag, and K$=14.07\pm0.06$ mag. 
The J band upper limit could be due to the drop of the UV-optical spectrum 
at  these wavelengths, but the 2MASS color index (H-K)$=1.56$   does  not support the contribution from any donor late type 
star 
\citep[see][]{Straizys2009}. If the 2MASS J band 
upper limit is attributed to a M5 star (for an orbital period of 1.69 h), the minimum distance is then 
400 pc. This is more compatible with the low interstellar absorption derived from the X-rays.

\begin{figure}
\includegraphics[width=6.cm,height=8.cm,angle=-90]{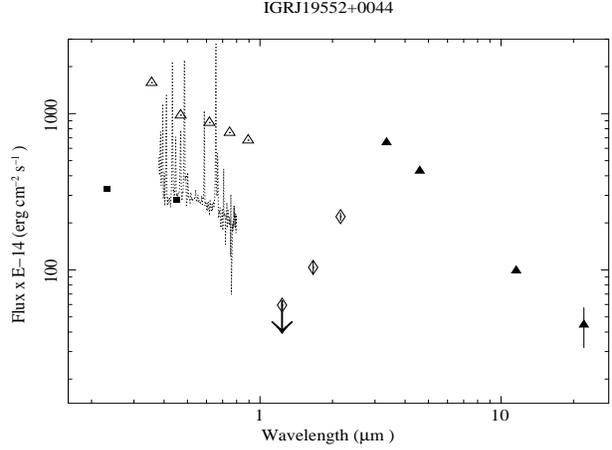}
\caption{The SED of \JC\ constructed with the \emph{XMM-Newton} UV and B band fluxes (filled squares), the optical spectrum by \citet{Masetti10} (dotted line), the SDSS photometry (empty triangles), the 2MASS nIR (empty diamonds) and WISE measures (filled triangles). No correction for extinction is applied. The
peculiar shape is discussed in the text.}
\label{fig:igr1955sed}

\end{figure}

\begin{figure}
\includegraphics[width=\columnwidth,height=6.5cm,angle=0]{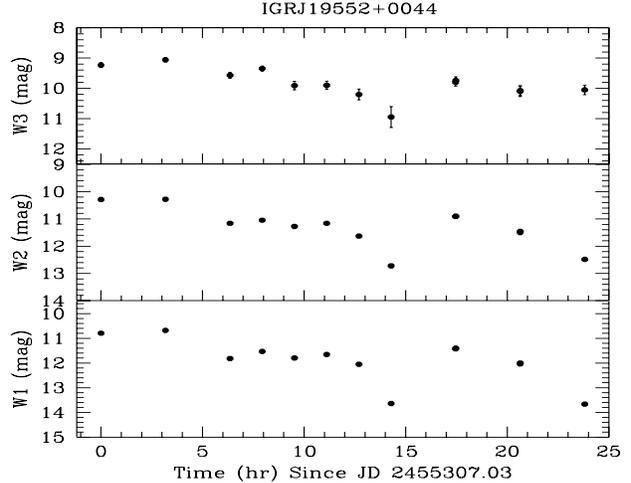}
\caption{The \emph{WISE} light curve of \JC\ constructed using the "single exposures" in the W1, W2 and W3 
bandpasses, showing large IR  variability}.
\label{fig:wiselc}
\end{figure}

\noindent The peculiar shape of the nIR spectrum is further confirmed by the surprising detection of \JC\ at IR 
wavelengths 
by the Wide-field Infrared Survey  Explorer  \emph{WISE} \citep{Wright10} in all the four bands, W1 (3.35$\mu$m), W2 
(4.6$\mu$m), W3 (11.6$\mu$m) and W4  (22.1$\mu$m) and catalogued as WISE\,J195512.47+004536.6. 
The corresponding magnitudes are $\rm m_{W1}=11.556\pm0.024$, $\rm m_{W2}=11.031\pm0.023$, $\rm 
m_{W3}=9.704\pm0.041$ and $\rm m_{W4}=8.506\pm0.31$. 
Most WDs detected by \emph{WISE} (the \emph{WISE} InfraRed Excesses around Degenerates, WIRED, survey 
project) are WD + Main Sequence binaries \citep{Debes11}. Infrared excess around WDs could  
also be due to debris  discs \citep{Debes12} or cyclotron emission in case of magnetic systems 
\citep{Harrison13}.
Therefore, we have combined the OM UV and B band photometry, the SDSS (u',g',r',i and z') and 
optical spectrum by \citet{Masetti10} with the 2MASS (J, H and K) and WISE (W1,W2, W3 and W4) photometry 
to  construct  the spectral energy distribution (SED),  although these measures are 
taken at different epochs.\footnote{The SDSS 
data were taken in Sept. 2004, the 2MASS data  in Aug. 1999, the WISE photometry in Apr. 2010 and the 
optical spectroscopy in 2009.} 
(see Fig.\ref{fig:igr1955sed}). The SED  is not corrected for extinction because, from the hydrogen column 
density of X-ray fits, it is at most $\rm E_{(B-V)}$=0.1.  
Nevertheless, the unusual shape of the SED is apparent,  with a pivot energy at 1.2\,$\mu$m, where at shorter 
wavelengths the spectrum is a power-law $\rm F_{\nu} \propto \nu^{\alpha}$, with $\alpha$=0.9 for the OM UV and B  band 
data, and $\alpha$=1.5 for the SDSS data. The rapid rise of the 
flux   beyond 1.2\,$\mu$m and the peak in the W1 (3.35$\mu$m) is very similar to that observed in the Polar EF\,Eri with 
\emph{WISE}  and \emph{Spitzer} 
\citep{Harrison13}. EF\,Eri is a short period (1.35 h) Polar believed to harbour a sub-stellar secondary star and 
which has 
been in a prolonged low state for a decade. The nIR and IR spectrum of EF\,ERi also does not give evidence of the 
secondary  star but has been modeled as due to cyclotron emission from  accretion  onto a 12.6\,MG  WD \citep[see][for more 
details]{Harrison13}. 
At these wavelengths the 
contribution comes from lower  cyclotron harmonics ($n$=1,2,3), which are expected to be optically thick.
The lack of nIR and IR spectroscopy does not allow a meaningful analysis of the spectrum of \JC\ with cyclotron models.
Also, since EF\,Eri is strongly variable (more than a magnitude) in the nIR and IR along 
its orbital period, we have then inspected the "single exposure" observations from \emph{WISE} 
survey. These were acquired along 1 d on Apr.\,21, 2010. Useful data are reported for the W1, W2 
and W3 bandpasses in Fig.\ref{fig:wiselc}, that show large  variations ($\Delta\,M>1$ mag). Unfortunately the data are too 
sparse to phase them at either 1.38 h or 1.69 h periods.

\noindent Hence, the two short periodicities, the X-ray spectral characteristics and the striking 
similarity of the nIR and 
IR properties of \JC\ with EF\,Eri favours a  magnetized CV accreting at a low accretion rate below the orbital period 
gap.


\section{Summary}

We have presented the first long X-ray observations of the 
three new hard X-ray sources \JA, \JB\ and \JC\ together with UV, optical, nIR and IR 
photometry.

\noindent All three sources are hard X-ray emitting CVs that show large amplitude variabilities
on timescales of minutes to hours. Two of them, \JA\ and \JC,  also show long (months to years) variabilities.
Their broad-band X-ray spectra present characteristics typical of magnetic CVs. We here summarize the main results:

\begin{itemize}
\item 
Swift\,J1907 is a long (20.82 h) orbital period CV. We detected a non-coherent variability at a
period of  $\sim$7200 s. It  is likely of transient nature, reflecting matter traveling 
from the secondary star onto the compact object. 
\JA\ reveals a spectrum closely resembling those of magnetic CVs, including a soft X-ray component
originating from a small fraction of the WD area.
A magnetic system is then favoured although we cannot give a proper classification.
The supra-solar abundance of nitrogen (about twice the solar value),  the inverted mass ratio (q$=1.3$) and the long orbital period 
could suggest that \JA\ underwent a TTMT phase.
\item 
IGR\,J1212 reveals a  X-ray variability at $\gtrsim$7.6 h that might  reflect the binary 
orbital period. No fast 
coherent pulses are detected though, but its spectral characteristics give support to a magnetic CV of the IP 
type. 
\item  
IGR\,J1955 shows two X-ray periods: 1.38 h and 1.69 h. Its X-ray spectrum indicates a system 
accreting at low 
rate similarly to the CVs below the orbital period gap. We have found \JC\ to display peculiar nIR and IR 
spectrum remarkably similar to that observed in the Polar EF\,Eri. We propose this system to be a pre-polar or an IP 
with the lowest degree of asynchronism, resembling the peculiar magnetic CV Paloma. 
\end{itemize}

\noindent Our study of newly discovered CV candidates has revealed new features, 
departing from canonical view of magnetic CVs. In general,  most of the hard X-ray discovered CVs are being found at 
relatively long orbital periods. This aspect deserves a  detailed investigation to understand the evolutionary status of 
these CVs.

\section*{Acknowledgments}
This work is based on observations obtained with
\XMM\ an ESA science mission with instruments and contributions
directly funded by  ESA Member States; with \Swift, a NASA science
mission with Italian participation; and \emph{Suzaku}, a collaborative mission between the space agencies of Japan (JAXA) and the USA (NASA).
This publication also makes use of data products from the Wide-field Infrared Survey Explorer,  which is a joint project of the University of California, Los Angeles, and the Jet Propulsion Laboratory/California Institute of Technology, funded by the National Aeronautics and Space Administration; the Two Micron All Sky Survey (2MASS), a joint project of the University
of Massachusetts and the Infrared Processing and Analysis Center (IPAC)/Caltech,
funded by NASA and the NSF; and the Sloan Digital Sky Survey (SDSS). 
\bibliographystyle{mn2e}
\bibliography{biblio}

\vfill\eject
\end{document}